# Bound states in the continuum in a two-dimensional $\mathcal{PT}$-symmetric system

YAROSLAV V. KARTASHOV,[1,2,*] CARLES MILIÁN,[1] VLADIMIR V. KONOTOP,[3] AND LLUIS TORNER[1,4]

[1]*ICFO-Institut de Ciencies Fotoniques, The Barcelona Institute of Science and Technology, 08860 Castelldefels (Barcelona), Spain*
[2]*Institute of Spectroscopy, Russian Academy of Sciences, Troitsk, Moscow, 108840, Russia*
[3]*Centro de Fisica Teórica e Computacional and Departamento de Física, Faculdade de Ciências, Universidade de Lisboa, Campo Grande 2, Edifício C8, Lisboa 1749-016, Portugal*
[4]*Universitat Politècnica de Catalunya, 08034, Barcelona, Spain*

We address a two-dimensional parity-time ($\mathcal{PT}$)-symmetric structure built as a chain of waveguides, where all waveguides except for the central one are conservative, while the central one is divided into two halves with gain and losses. We show that such a system admits bound states in the continuum (BICs) whose properties vary drastically with the orientation of the line separating amplifying and absorbing domains, which sets the direction of internal energy flow. When the flow is perpendicular to the chain of the waveguides, narrow BICs emerge when the standard defect mode, which is initially located in the finite gap, collides with another mode in a standard symmetry breaking scenario and its propagation constant enters the continuous spectrum upon increase of the strength of gain/losses. In contrast, when the energy flow is parallel to the chain of the waveguides, the symmetry gets broken even for a small strength of the gain/losses. In that case, the most rapidly growing mode emerges inside the continuous spectrum and realizes a weakly localized BIC. All BICs found here are the most rapidly growing modes, therefore they can be excited from noisy inputs and, importantly, should dominate the beam dynamics in experiments.

Bound states in the continuum (BICs) are modes encountered in various potentials that share one remarkable and somewhat counter-intuitive property: despite their eigenvalues fall in part or in total into the continuous domain of the spectrum, the modes remain radiationless and thus spatially localized. Typically, BICs emerge under conditions where a suitable destructive interference leads to the complete cancellation of the outgoing radiation. Thus, BICs appear at the proper values of the parameter space where such a cancellation is possible (see the review [1]). The concept of a BIC was put forward originally by von Neumann and Wigner in [2] under conditions where synthetic potentials were engineered to support them. Several decades later, the concept was revisited and extended to coupled resonances [3,4]. Today different powerful theoretical approaches to construct potentials supporting BICs are known. They include ideas inspired in the techniques developed for super-symmetric quantum mechanics [5,6] or for engineering reflectionless potentials [7], as well las formalisms based on Darboux transformation [8], separability of Hamiltonians [9], or coupled spinor systems [10]. Experimental observations of BICs were reported in acoustics early [11] and more recently in diverse optical settings [12-23]. Optical BICs may form, in particular, in periodic photonic crystal slabs [12,13], arrays and photonic crystals with side-coupled defects [15-17], in waveguide arrays with engineered coupling strengths [18,19], parallel arrays of dielectric cylinders [20], in fully anisotropic media [21], and in lasing systems [22], to name just a few settings. In most cases studied to date, the potentials supporting the BICs are conservative and are formed only by a suitable refractive index modulation.

So-called $\mathcal{PT}$-symmetric potentials are complex, with real and imaginary parts, but they however feature a completely real modal spectrum due to a specially-designed, balanced profile of gain and losses set by the imaginary part of the potential, provided that the strength of gain/losses falls below a certain critical level, known as the $\mathcal{PT}$-symmetry breaking threshold. The properties of the usual guided modes in $\mathcal{PT}$-symmetric potentials are well-known (see, e.g., the review papers [24-26]), and experiments are possible after the landmark demonstration reported in [27]. A distinctive property of a $\mathcal{PT}$-symmetric potential is that the shapes of the guided modes and the beam propagation are strongly affected by the imaginary part of the potential, thus offering a new degree of freedom to manipulate the modal structure.

Thus, a question arises about whether by tuning gain/losses one can generate BICs in such complex potentials. They were encountered indeed in one-dimensional and discrete $\mathcal{PT}$-symmetric waveguide arrays with engineered coupling constants [28-32] and defect modes similar to BICs were observed in discrete mesh lattices [33]. However, higher-dimensional BICs in $\mathcal{PT}$-symmetric potentials have not been addressed so far.

In this Letter, we show that BICs may form in a two-dimensional continuous $\mathcal{PT}$-symmetric system formed as a chain of waveguides, where all channels are conservative except for the central one, which has both amplifying and absorbing domains. We reveal that the properties of the BICs emerging upon the increase of the strength of gain/losses in the central waveguide strongly depend on the orientation of the internal energy flow in the dissipative channel with respect to the waveguide chain. We describe two different types of BICs both occurring above the $\mathcal{PT}$-symmetry breaking threshold. One of them appears when a defect mode from the gap enters the continuum. The other one emerges directly inside the continuous spectrum. Note that here we refer to BICs that exhibit complex eigenvalues, with real and imaginary parts and whose real parts belong to the continuous part of the spectrum.

The propagation of light in our system is described by the two-dimensional Schrödinger equation for the dimensionless field amplitude $q$:

$$i\frac{\partial q}{\partial z} = -\frac{1}{2}\left(\frac{\partial^2 q}{\partial x^2} + \frac{\partial^2 q}{\partial y^2}\right) - \mathcal{R}(x,y)q, \quad (1)$$

where the transverse coordinates $x, y$ are scaled to the characteristic width $w_0$ and the propagation distance $z$ is scaled to the diffraction length $2\pi n w_0^2/\lambda$. The $\mathcal{PT}$-symmetric potential $\mathcal{R} = p_{\rm re}\mathcal{R}_{\rm re} + ip_{\rm im}\mathcal{R}_{\rm im}$ contains real and imaginary parts with depths $p_{\rm re}$ and $p_{\rm im}$, respectively. The real part of the potential models an infinite array of single-mode Gaussian waveguides $\mathcal{Q} = \exp[-(x^2+y^2)/a^2]$ with width $a$, separated by a distance $d$ along the $x$-axis: $\mathcal{R}_{\rm re} = \sum_m \mathcal{Q}(x-md,y)$ (see insets in Fig. 1). We assume that gain and losses are present only in the central waveguide and that the imaginary part of the potential is given by $\mathcal{R}_{\rm im} = (x\cos\alpha + y\sin\alpha)\mathcal{Q}$, where the angle $\alpha$ determines the orientation of the amplifying and absorbing domains in the central waveguide with respect to the waveguide chain. The potential satisfies the condition $\mathcal{R}(x,y) = \mathcal{PT}[\mathcal{R}(x,y)] = \mathcal{R}^*(-x,-y)$. Eigenmodes of $\mathcal{PT}$-symmetric system are characterized by the presence of the internal currents from the amplifying into the absorbing domains [26]. When $\alpha = \pi/2$, the current is perpendicular to the waveguide chain [inset in Fig. 1(a)], while for $\alpha = 0$ it is parallel to the waveguide chain [inset in Fig. 1(c)]. The properties of the two-dimensional BICs differ substantially for different current directions (angles $\alpha$), as shown below.

When $p_{\rm im} = 0$ the eigenmodes of the waveguide chain are Bloch waves $q = u(x,y)\exp(ib_m z + ikx)$, where $b_m$ is the propagation constant, $k$ is the Bloch wavenumber in the first Brillouin zone, and $m$ is the number of the allowed band. The functions $u(x+d,y) = u(x,y)$ are periodic in $x$ and depend on $b_m$ and $k$. For $b_m > 0$ the functions $u(x,y)$ decay exponentially at $y \to \pm\infty$, i.e. far from the waveguide chain (see the upper gray region in Fig. 1, where Bloch modes are localized in $y$ and have intensity maxima coinciding with the waveguides), but they are delocalized along $y$ for $b_m < 0$ (see the lower gray region in Fig. 1), thus $b_m = 0$ gives the upper edge of the continuum. In the numerical simulations below we set $p_{\rm re} = 6$, $a = 0.5$, $d = 1.8$. The continuous spectrum of an array with such parameters at $p_{\rm im} = 0$ is shown with gray color in Fig. 1. There is a semi-infinite forbidden gap at $b > 1.523$ and one finite gap at $0 < b < 0.696$ (it is shown white). Here we address the impact of the parameters $\alpha$ and $p_{\rm im}$ on BICs emerging in the system, in the presence of gain/losses in the central channel.

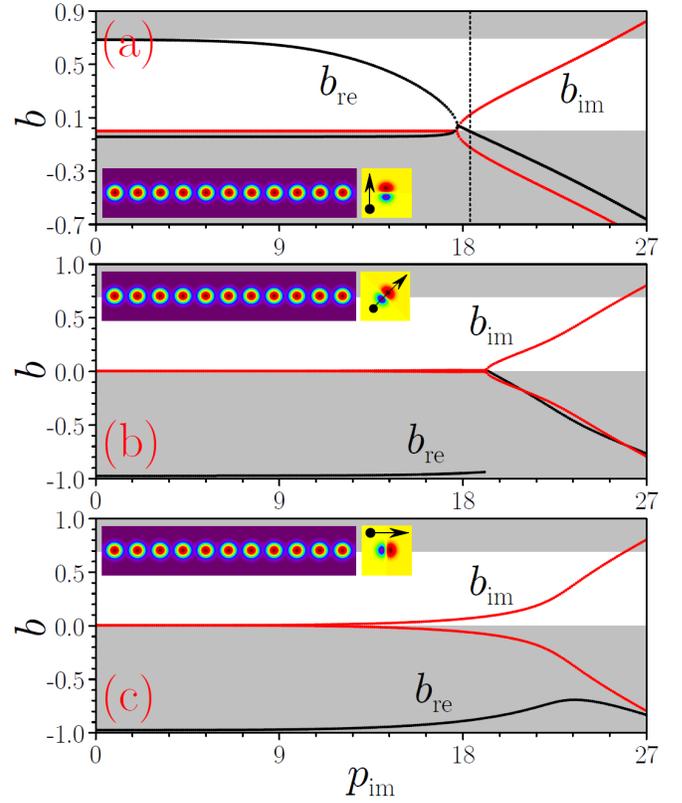

Fig. 1. (Color online) Real $b_{\rm re}$ (black dots) and imaginary $b_{\rm im}$ (red dots) parts of the propagation constant for BICs or modes, whose collision gives rise to BICs, versus $p_{\rm im}$ at $\alpha = \pi/2$ (a), $\alpha = \pi/4$ (b), and $\alpha = 0$ (c). Gray regions indicate the continuous spectrum. The vertical dashed line in (a) indicates $p_{\rm BIC}$. Insets show the real part of the refractive index in the array and the gain/loss distribution in the central channel. Arrows indicate the directions of the currents for each arrangement of gain/losses.

When $p_{\rm im} \neq 0$ the central waveguide can be considered as a defect of the array because the coupling constant between the central waveguide and its neighbors changes [34]. At $\alpha = \pi/2$, when in addition to usual $\mathcal{PT}$-symmetry the condition $\mathcal{R}(x,y) = \mathcal{R}^*(x,-y)$ holds, the defect with small $p_{\rm im}$ does not break the $\mathcal{PT}$-symmetry. Rather, it leads to the appearance of a localized defect mode $q = u(x,y)\exp(ibz)$ with real propagation constant $b$ in the first finite gap. Figures 1(a) and 2(a) show, respectively, the propagation constant and the integral width $w = \iint (x^2+y^2)^{1/2}|u|^2 dxdy$ of such mode as functions of $p_{\rm im}$ (we use the normalization $\iint |u|^2 dxdy = 1$). Representative profiles of the defect mode are shown in Fig. 3 (first row, left). The mode becomes more localized with increase of $p_{\rm im}$, while its propagation constant (eigenvalue) gradually approaches the lower gap edge. In the exceptional point $p_{\rm im} = p_{\rm EP}$ the defect mode coalesces inside the gap with another mode emerging from the continuous spectrum, as shown in Fig. 1(a). This is the most typical scenario of $\mathcal{PT}$-symmetry breaking, which leads to the appearance of modes with complex-conjugate eigenvalues $b = b_{\rm re} \pm ib_{\rm im}$, whose real parts $b_{\rm re}$ enter the continuous spectrum upon increase of gain and losses, at some $p_{\rm im} = p_{\rm BIC} > p_{\rm EP}$. Then, there appear two BICs, one of which grows and the other one decays upon propagation (see Fig. 3, left column, rows 2 and 3). The width of such BICs slowly decreases with $p_{\rm im}$ [Fig. 2(a)]. It should be stressed that we refer to these states as BICs on account of the fact that real parts of their eigenvalue fall into the continuous spectrum, even though the whole eigenvalue has also an imaginary part. Thus, ac-cording to the classification introduced in [28], such states are type-I BICs. This scenario with the defect mode turning into a BIC and a clear transition from the real to the complex spectrum was encountered only for one particular angle $\alpha=\pi/2$, at which the internal current is perpendicular to the waveguide chain. BICs appearing due to this mechanism are typically well-localized. Note that the continuous character of Eq. (1) and the presence of a finite gap at $p_{\rm im}=0$ are crucial ingredients for the existence of such states.

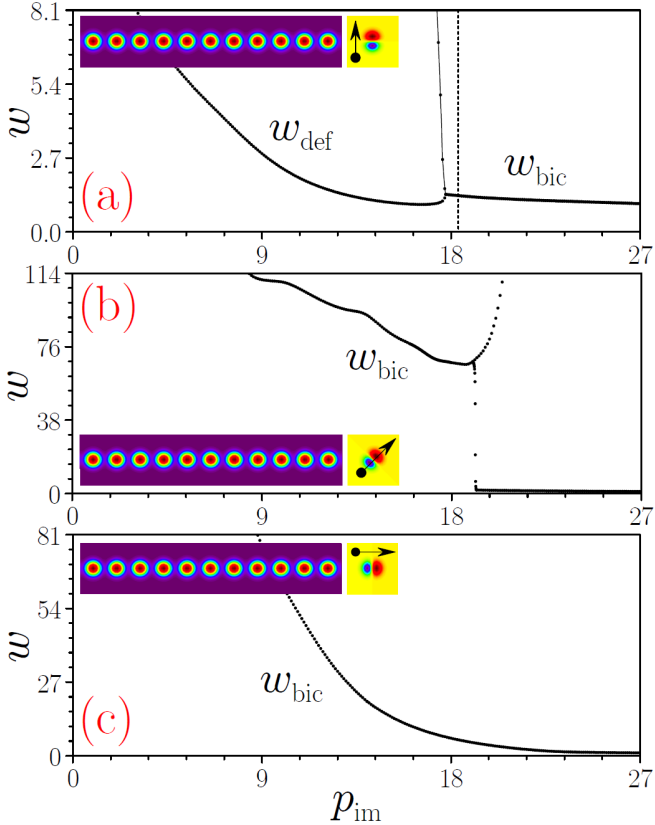

Fig. 2. (Color online) Integral width of BICs or modes, whose collision gives rise to BICs, versus $p_{\rm im}$ at $\alpha=\pi/2$ (a), $\alpha=\pi/4$ (b), and $\alpha=0$ (c). The meaning of insets is the same as in Fig. 1. Note the differences in scales of the $w$-axes.

The picture changes *qualitatively* for smaller angles. Even though the defect mode discussed above still exists for $\alpha<\pi/2$ and that a decrease of $\alpha$ results in a shift of $p_{\rm EP}$, the $\mathcal{PT}$ symmetry breaking may occur not for the gap states, but for modes in the continuous spectrum. Such scenario of the symmetry breaking was recently found numerically [35] and explained analytically [36] for some classes of one-dimensional non-Hermitian optical potentials. In our case such transition occurs already at very small values of the imaginary part of the potential, as visible in Fig. 1(b) where we show the eigenvalues with maximal imaginary part at $\alpha=\pi/4$. The real parts of the corresponding eigenvalues fall deeply inside the continuous spectrum ($b_{\rm re}\sim-1$), while the imaginary parts are so small that they are not even visible in the figure. The corresponding modes represent BICs of the second type mentioned above. Because they bifurcate from the modes of the continuous spectrum, they are weakly localized. We found that the width of such modes changes with $p_{\rm im}$ nonmonotonically [Fig. 2(b)]. However, when $p_{\rm im}$ reaches a certain threshold level ($p_{\rm im}\sim19.1$), another pair of well-localized modes with com- plex-conjugate propagation constants appear. Such pair represents well-localized BICs similar to those encountered at $\alpha=\pi/2$. Since these last modes become dominating in the evolution due to their much larger imaginary parts $b_{\rm im}$, we switch to their parameters in Fig. 1(b) and 2(b) at $p_{\rm im}>19.1$. In other words, numerically we detect the mode with the largest amplification, which explains the discontinuity of the $b_{\rm re}$ curves in Fig. 1(b). When a pair of new modes emerge at $p_{\rm im}>19.1$, their width initially rapidly decreases, but then changes much slower for large $p_{\rm im}$ values [see lower branch in Fig. 2(b)].

When $\alpha=0$ and the current is parallel to the waveguide chain, a new behavior emerges. Now $\mathcal{R}(x,y)=\mathcal{R}^*(-x,y)$ and BICs again emerge inside the continuous spectrum at very small $p_{\rm im}$ [Fig. 1(c)]. Their localization progressively and monotonically increases with $p_{\rm im}$ [Fig. 2(c)]. By varying $p_{\rm im}$ and following the BIC with the largest imaginary part of the eigenvalue, i.e. with the largest increment, at $\alpha=0$ we did not encounter any sharp variations in the mode width, like those shown in Fig. 2(b) for $\alpha=\pi/4$. This means that the BIC that emerges from the continuum at the point of $\mathcal{PT}$-symmetry breaking exhibits an amplification larger than that of the defect modes. An example of a BIC for $\alpha=0$ at the intermediate $p_{\rm im}$ value is shown in Fig. 3 (first row, right). Such BIC is characterized by long, but localized tails. For larger values of $p_{\rm im}$ such states become more confined in the uniform medium than inside the lattice, as illustrated in Fig. 3, right column, rows 2 and 3. The real part of the propagation constant of the BIC exhibits a nonmonotonic behavior, but always stays within the continuous spectrum [see Fig. 1(c)]. The BICs of the second type, found for $\alpha=0$ and $\alpha=\pi/4$, exhibit field amplitudes that peak in between the waveguide cores. This feature is reminiscent of the Bloch modes ($p_{\rm im}=0$) that arise in the lowest allowed band of the system from where the BICs emerge.

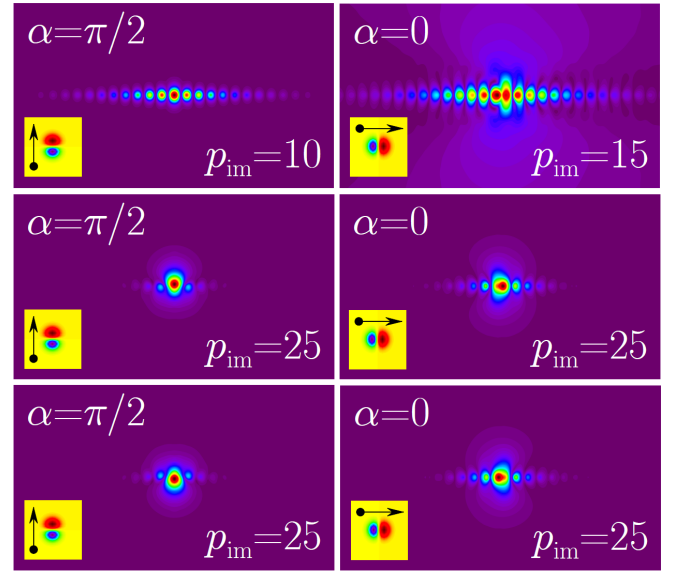

Fig. 3. (Color online) Localized modes supported by a waveguide chain with $\alpha=\pi/2$ (left column) and $\alpha=0$ (right column) at different $p_{\rm im}$ values. All these modes are BICs expect for the left mode in the top row that represents usual defect mode. Modes in the second row have $b_{\rm im}>0$ and decay, while modes in the third row have $b_{\rm im}<0$ and grow. The insets show the distribution of gain/losses in the central waveguide and the current directions. Modes are shown within $x\in[-24,+24]$ and $y\in[-16,+16]$ window.

It is important to emphasize that all BICs obtained here exist in the regime of broken $\mathcal{PT}$ symmetry and represent the modes of the system with fastest exponential growth (or decay for states with complex-conjugate eigenvalues). This is relevant for their experimental realization and exploitation, because for any input overlapping with such BICs one should observe amplification exactly in the form of the BIC states. This feature is illustrated in Fig. 4, which shows the field modulus distributions at different distances obtained by solving Eq. (1) with noisy inputs and for two different angles $\alpha=\pi/2$ and $\alpha=0$. In both cases BICs determine the output distribution after relatively short propagation distances (note that the dynamically determined propagation constants of these states have $b_{\rm re}<0$).

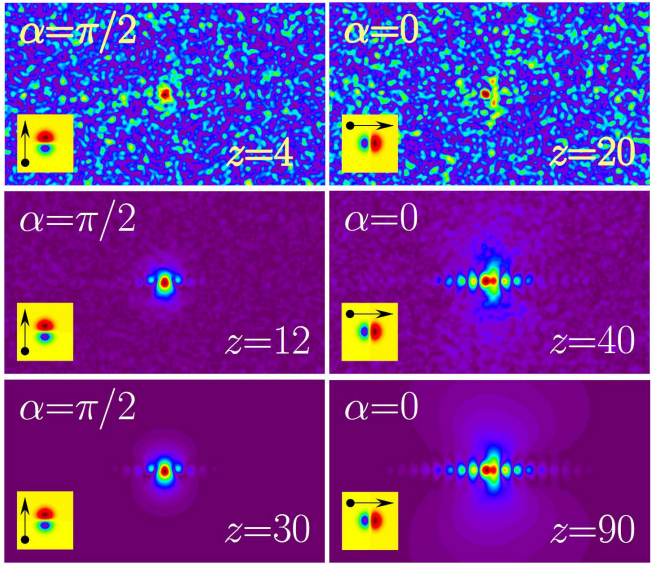

Fig. 4. (Color online) Dynamical generation of BICs starting from noisy field distributions for $\alpha=\pi/2$ (left column) and $\alpha=0$ (right column) at $p_{\rm im}=20$.

Summarizing, we addressed a specially-designed, two-dimensional $\mathcal{PT}$-symmetric structure that was shown to support BICs whose propagation constants, symmetry, and widths strongly depend on the orientation of the energy current inside the only dissipative channel with respect to a chain of conservative waveguides. Two different types of BICs were encountered. BICs of the first type are strongly localized and represent the continuation of the defect modes into continuum. BICs of the second type bifurcate from a mode in the continuum and are weakly localized for moderate gain and losses. The described BICs realize the modes with fastest exponential growth in the system.

We acknowledge support from the Severo Ochoa program (SEV-2015-0522) of the Government of Spain, from Fundacio Cellex, Fundació Mir-Puig, Generalitat de Catalunya, and CERCA.